# Ultrasound field measurement using a binary lens

G.T. Clement[1], H. Nomura[2], and T. Kamakura[2]

1. Department of Biomedical Engineering, Cleveland Clinic, Cleveland, OH USA
2. Graduate School of Informatics and Engineering, The University of Electro-Communications, Chofu-Shi, Tokyo, Japan

Abstract-

Field characterization methods using a scattering target in the absence of a point-like receiver have been well described in which scattering is recorded by a relatively large receiver located outside the field of measurement. Unfortunately, such methods are prone to artifacts due to averaging across the receiver surface. To avoid this problem while simultaneously increasing the gain of a received signal, the present study introduces a binary plate lens designed to focus spherically-spreading waves onto a planar region having a nearly-uniform phase proportional to that of the target location. The lens is similar to a zone plate, but modified to produce a bi-convex-like behavior, such that it focuses both planar and spherically spreading waves. A measurement device suitable for characterizing narrowband ultrasound signals in air is designed around this lens by coupling it to a target and planar receiver. A prototype device is constructed and used to characterize the field of a highly-focused 400 kHz air transducer along 2 radial lines. Comparison of the measurements with numeric predictions formed from nonlinear acoustic simulation showed good relative pressure correlation, with mean differences of 10% and 12% over center 3dB FWHM drop and 12% and 17% over 6dB.

I. Introduction

Size, geometry, sensitivity, and range of linearity all play a significant role in the proper choice of an ultrasound receiver [1]. Such considerations are particularly important in transducer field characterization, where the receiver must be sufficiently small to allow assumptions of uniform amplitude and phase over its active surface [2]. This can be problematic for highly focused fields [3] and in certain applications at relatively high frequencies [4], where small yet adequately-sensitive receivers can be costly, or even altogether unavailable. For example, we found to be the case in our ongoing investigations [5] utilizing highly focused air ultrasound at frequencies at or above 400 kHz.

Optical approaches have been proposed to meet the specific challenges for this illustrative case of high frequency and high pressure in air, which lacks suitably-small microphones. Tomographic field reconstructions have been demonstrated [6]–[8], however these were performed for cases involving of low pressure variation, such that higher light diffraction orders could be neglected. Such conditions are not valied at higher intensities [9], [10], making accurate field reconstruction problematic.

Another well-established solution is the introduction of a small point-like (or line-like [4], [11]) target into the field[12], [13], thus inducing a spherical (cylindrical) wave that can be recorded by a larger receiver placed outside of the field. Precise relative measurement of the complex pressure generally requires that the target be smaller than source wavelength [2], and is further restricted to half this wavelength when the oncoming signal - or a portion of its wavevector components- are sharply angled relative to the target cross-section [14]. Receivers, on the other hand, can be arbitrarily large, as long as they are geometrically matched and normally aligned to the spreading wavefront, or otherwise capable of detecting phase differences over the measurement surface. For amplitude measurements, receivers [15] or scatterers larger than a wavelength can be used with a deconvolution approach [16], however such methods are inadequate for applications where phase information is critical.

In the present work, we investigate an alternative target scattering method designed around an acoustic binary lens. It will be shown that a zone-plate-type lens can focus a monochromatic spherically-scattered wavefront onto a small region having uniform phase over its cross section. A receiver whose active surface is confined to this region would then be expected to record an amplitude and relative phase corresponding to that of the initial scattering location. This preliminary work provides the basis of the approach, and points out some interesting relevant properties of this lens. To verify the method for our particular application in airborne ultrasound, a prototype system is designed and constructed. This system is then used to perform basic characterization of a 400 kHz High Intensity Focused Ultrasound (HIFU) air transducer.

## II. Binary Lens

The measurement approach is based around a lens design comprising a variation on the classic Fresnel binary zone plate [17]. Zone plates consist of alternating transmitting and opaque rings centered about an axis and spaced with inner ($n$, odd) and outer ($n$, even) edges located at

$$r_n = \sqrt{\left(\sqrt{r_0^2 + L^2} + n\frac{\lambda}{2}\right)^2 - L^2}, \tag{1}$$

where $\lambda$ is the wavelength and $r_0$ is centermost edge of arbitrary length. A plane wave normally incident upon the plate undergoes diffraction such that constructive interference occurs at the desired focal length $L$, on the axis of symmetry. Zone plates have been used for focusing [18], [19] and incorporated into transducer designs [20], [21], including air-coupled designs[15], as have the closely-related Fresnel Lens [22], which shifts phase alternately by 0 and π radians along rings defined by Eq (1). In one sense, the zone plate may be viewed as further simplification of this process, keeping only the 0 radian "shifts" of the Fresnel lens and blocking out all other parts of the field.

On the other hand, optimal superposition at a given focal point requires that all incoming waves arrive precisely in phase, suggesting an ideal binary lens would consist of a series of infinitesimal circular slits located in a plane at radial distances

$$r_n = \sqrt{\left(\sqrt{r_0^2 + L^2} + n\lambda\right)^2 - L^2} \tag{2}$$

equal to the mean radial distance between the open rings of (1). Yet clearly, in consideration of total signal strength, the slits must be of finite size and a tradeoff must occur between the magnitude of the slit area and the phase mismatch this finite area produces (Figure 1). Indeed, as a given slit centered about each $r_n$ is expanded in radius, it will continue to focus an incident wave with stronger overall signal strength up to the point where ± π/2 rad in phase variation is transmitted over the rings, precisely the point the lens becomes identical to the classic binary zone plate.

Returning to the ideal case of infinitesimal slits, it may be readily verified that a spherical wave expanding from the lens focal point will have uniform phase at the slit locations. This implies not only a planar wave, but also this incident spherical wave will be focused by the lens (Figure 2). Such behavior is attune to that of a bi-convex lens [17] and is one key characteristic exploited in the present method. As with planar focusing, a tradeoff occurs between the slit size and overall focusing ability. It will be shown in Sec III.A, that as ring diameter is increased, the focal strength of the bi-convex focus increases until ± π/4 rad phase variation is passed through each ring. Further increasing ring size beyond this point causes a decrease in peak signal, and the bi-convex property disappears altogether when ± π/2 rad phase variation is passed, *i.e.* when dimensions become identical to the zone plate. It is noted that values $r_n$ given by (2) will be multiples of any higher harmonic of the focusing wavelength of the lens. It would then be expected that - at least in the ideal case - harmonics passing through the lens would also be focused. However, it is further noted that rings of optimal dimension for the fundamental (i.e. those permitting ± π/4 phase variation) represent the *worst case* for the second harmonic, which passes ± π/2 phase through the same rings. This effect and the potential to use the lens design for imaging harmonics of the selected fundamental is examined in the simulation study (III) and experimental measurements (IV.*B*).

For use as a measurement tool, a point-like target is introduced at one focal point of the binary lens, which focuses scattering from the target onto the opposing focal plane. Measurement of the complex pressure necessitates the use of a target smaller than the source wavelength [2]. However, in the present method, this target is ideally no larger than ½-wavelength across in any direction. In this case, the focused wave will exhibit a nearly-uniform phase across the primary peak in the focal plane (Figure 3). This property, which is examined as part of a simulation study below, facilitates the placement of a circular planar receiver that has the same area as the focal cross-section. If such a target/lens/receiver configuration is rigidly fixed in a way that is minimally-invasive to the field, the resulting system produces a signal at the receiver proportional in phase to that of the target location. This ideal configuration is depicted in Figure 4. It is noted that a target having characteristic dimensions greater than ½-wavelength – but no greater than the source wavelength - might also be used, but with the consequences of introducing more complex directional scattering and thereby potentially introducing measurement error. Such a case, specific to our verification experiment, is considered in Section III.c.

## III Simulations

### A. Lens

A series of simulations were executed to aid in the lens design as well as the design of the overall system. Scattered fields incident upon a lens surface were calculated using a discrete approximation to the Rayleigh-Sommerfeld diffraction integral [23]. While there is no expectation that the field to be characterized is linear, the small target size and subsequent divergence of the scattered field compounded with at least a quadratic increase in absorption with frequency [24], [25] that readily justifies a linear approximation of the scattered wave.

The lens itself was treated as a one-dimensional (radial) diffraction grating and the transmitted field was projected throughout the focal region using the angular spectrum approach under the condition of axial symmetry [14], [26]. Assuming an incident continuous 400 kHz signal in air, fields as a function of their axial and radial dimensions were calculated out to a distance of 400 mm past the lens while keeping the outer diameter of the lens fixed at 60 mm.

These simulations were first used to examine focusing ability as a function of focal length and the degree of phase variation passed through the lens. For comparison, fields from both a plane wave and a point source were examined. In addition to the fundamental frequency, effects of the lens on the second (800 kHz) and third (1200 kHz) harmonics were also examined.

Focal distances ranging from L=0.5 mm to L=300 mm were calculated at 1 mm intervals. These computations were repeated for different ring openings, each expanded about a central radius given by (2), and whose radial thicknesses were determined by the phase range they permitted: That is, the maximum possible phase difference between any two positions located on any open ring, due to a spherically-spreading wave at a focal point. Ranges included ±π rad (open aperture), ±π/2 rad (*ie*. the zone plate given by (1)), ±π/4 rad, ± π/8 rad, and ± π/16 rad.

*B. Parameter study*

The optimal focal distance and optimal receiver diameter were determined by simulation, assuming point source excitation at the opposing focal point. In all cases, phase uniformity was observed within the first radial peak of the focal plane (Figure 3) and it was concluded that relative pressure measurement could be achieved using a receiver aligned parallel to the focal plane whose diameter did not exceed this first peak. Using the criteria of maximum receiver gain, results were evaluated to determine optimal focal distance, annulus dimension, and receiver size.

At each focal plane the receiver diameter was defined as the maximum diameter containing no more than 0.26 rad of phase variation (an arbitrarily-selected cutoff). This diameter covers the central flat region illustrated in Fig. 3b. The relatively small phase variation over this area estimated to contribute to under 2% overall relative phase error. Receiver sensitivity was assumed linearly proportional to both the surface area and the mean pressure over its active surface so that, although focal gain was found to decrease with focal distance (Figure 5), longer focal distances contained wider-diameters of uniform phase, and hence larger receivers.

As summarized in Figure 6 for the relevant case of a 60-mm-diameter lens and a 400 kHz signal, optimal placement was found to occur at focal length of approximately 133 mm when using a lens allowing ±π/4 rad slit openings. This was closely followed by a ±π/8 rad opening at the same focal distance. The corresponding optimal receiver diameter was 5.1 mm for the ±π/4 rad lens (Figure 3) corresponding to a 3.9X peak signal gain relative to the pressure at the lens surface. For the ±π/8 rad lens at the same distance, the optimal diameter was 5.8 mm and the expected relative peak gain 2.9X. Conversely, the classic zone plate (±π/2 rad) was observed to have the lowest sensitivity over the range of slit diameters studied.

*C. Target*

To approximate point-like scattering, a target would ideally consist of an object whose characteristic dimensions are much smaller than the imaging wavelength. Practical limitations detailed below in IV.1, however, limited the present target size to dimensions on the order of half the imaging wavelength (ka~2), prompting an estimation of scattering effects. Calculations were performed to determine the total pressure, $p = p_0 + p_s$, and particle velocity $\mathbf{u}(\mathbf{r}) = \mathbf{u}_0(\mathbf{r}) + \mathbf{u}_s(\mathbf{r})$ in terms of an incident ($_0$) and scattered ($_s$) wave. The relevant target (a fragment of steel wire) was assumed to approximate a finite rigid cylinder of radius $a$ and height $Z$, where the z-axis is the central axis of the lens. The total velocity normal to the surface was accordingly set to zero, such that in cylindrical polar coordinates ($r, \phi, z$), $u_{s_r} = -u_{0_r}$ on the body and $u_{s_z} = -u_{0_z}$ on the caps of the cylinder. Further approximating the incident wave in the region of the target as planar and propagating normal to the z-axis, the incident velocity normal to the surface could be expressed as

$$u_{0_r}(a,\phi,z) = \frac{i|p_0|}{\rho c} e^{ik_0 \cos(\phi)a} \cos\phi; \qquad |z| \leq \pm Z/2 \qquad (3)$$
$$0; \qquad otherwise$$

where $c$ is the sound speed in air, and $\rho$ the air density.

Scattered pressure on the boundary was determined by first writing the normal component of the scattered velocity on the surface in its Fourier integral form with respect to $z$, and by expanding remaining terms in general series [27]:

$$u_{s_r}(a,\phi,z) = k_r \sum_{m=-\infty}^{\infty} e^{im\phi} \int_{-\infty}^{\infty} A_m(k_z) e^{ik_z z} \left[ H^{(1)}_{m-1}(k_r a) - H^{(1)}_{m+1}(k_r a) \right] dk_z, \qquad (4)$$

where $H^{(1)}_m$ denotes an $m^{th}$ order Hankel function of the first kind. In this case, to satisfy the Helmholtz equation in polar coordinates, $k_r = \sqrt{k_0^2 - k_z^2}$. Substituting $u_{s_r} = -u_{0_r}$, and utilizing the orthogonality of the expansion, transformation yielded the expansion terms

$$A_m(k_z) = \frac{-2k_0}{\pi k_r} \frac{\sum_{m=-\infty}^{\infty} \int_{-\infty}^{\infty} u_{0r}(a,\phi,z) e^{-im\phi} e^{-ik_z z} dz}{H^{(1)}_{m+1}(k_r a) - H^{(1)}_{m-1}(k_r a)}, \qquad (5)$$

which could be used to provide the pressure at points on and exterior to the boundary:

$$p_s(r,\phi,z) = \sum_{m=0}^{\infty} e^{im\phi} \int_{-\infty}^{\infty} A_m(k_z) H^{(1)}_m(k_r r) e^{ik_z z} dk_z. \qquad (6)$$

Potential effect of a thread used to mount the target (ka~0.5) was similarly estimated by total contribution of the thread when located in the focal plane of the air HIFU transducer to be studied.

Based on the parameter study (III.2), scattering profiles were examined 133-mm from the scattering location over a 60-mm-diameter surface corresponding to the location and planned dimensions of the lens. It was determined that the integral effect of the thread would introduce a pressure error as large as 17% at the upper/lower periphery of the lens and as large as 33% at the center of the lens. Scattering profiles from these calculations about $z = 0$ mm and $z = 60$ mm are provided in Figure 7.

*D. Nonlinear field simulation*

To estimate system behavior under the approximate operating conditions of a focused transducer in air, as well as to simulate the transducer pressure field for comparison with experimental measurement, a nonlinear model was implemented based on a wavevector-frequency domain solution to the Westervelt equation. This approach, described in detail in [28], was specifically developed for its ability to accurately model both omnidirectional nonlinear waves as well as cases of frequency-dependent attenuation. As such, the approach provides an appropriate model equation [29] and methodology [30] for focused transducers.

Simulation of the 30-cycle burst used in the study was considered over the temporal frequency range from DC to 12 MHz (30 harmonics) and maximum radial wavenumber, $k_\rho = 2.2 \times 10^5 \frac{\text{rad}}{\text{m}}$, using a nonlinearity coefficient of $\beta = 1.2$ (diatomic gas), an ambient sound speed of c = 345 m/s (22° C, 1 atm), sea-level air density 1.2 kg/m². Absorption as a function of frequency, $f$, was set according to

$$\alpha(f) = 1.6 \times 10^{-10} \tfrac{\text{Np·s}^2}{\text{m}} f^2 - 9.8 \times 10^{-6} \tfrac{\text{Np·s}}{\text{m}} f + 4.7 \tfrac{\text{Np}}{\text{m}}, \tag{7}$$

derived from a quadratic fit of absorption data provided in [24]. Though the number of harmonics considered is relatively low, and thereby subject to distortion caused by the Gibbs phenomenon [31], such instabilities can be treated by artificial inflation of the absorption term for higher frequencies in combination with post-numerical recovery of higher components [32], which was implemented here.

Whereas the goal of the study was to measure the *relative* phase and pressure amplitude of the 400 kHz component of the field, spatial behavior of these relative values is pressure dependent, thus requiring knowledge of the source velocity or pressure *a priori*. However, justification for modeling the relative field from simulation of a just a rough estimation of the field strength may be argued based on the observation that spatial field characteristics tend to vary slowly as a function of input pressure. In the present case, focal pressures on the order of 160 dB were expected (described in Sec. IV.B), so that the field was simulated based on this estimated peak sound pressure and then normalized. Two fields were then modeled at significantly lower (155 dB) and higher (165 dB) than expected peak pressures, revealing a percent difference of 7%, as determined by

$$d_{400\text{kHz}} = 100 \frac{\sum_{m,n} \left\| |p_{155\text{dB}}(\mathbf{r}_{m,n})| - |p_{165\text{dB}}(\mathbf{r}_{m,n})| \right\|}{\sum_{m,n} \left\| |p_{155\text{dB}}(\mathbf{r}_{m,n})| + |p_{165\text{dB}}(\mathbf{r}_{m,n})| \right\| / 2}, \tag{8}$$

using the relevant 400 kHz component of the field and thus indicating the utility of the simulation over this range, to with within this error.

Scattered pressure on the boundary of the target was determined by its placement in the field and assuming hard boundary conditions using (6). This scattered surface pressure was then modeled as a source and propagated toward the lens along its axis of symmetry. It is noted that beyond a distance of approximately 10 wavelengths, the scattered wave was observed to exhibit linear-like behavior due to the combination of divergence and strong frequency-dependent attenuation. The lens was modeled as a perfect grating, where the forward components of the passing wave were propagated to the receiver plane, located 130 mm from the lens. The simulated amplitude was observed to have a reduced focal gain as compared with the ideal case, but similar relative shape (Figure 3a). Small variations (<0.1 rad) in the phase over the receiver surface (Figure 3b) due primarily to the finite size of the target were also observed.

## IV Laboratory Verification

### A. System Construction

Based on the parameter study (III.3) as well as the diameter of the receiver (R) available for the system (5 mm), a lens permitting a phase range of ± π/4 (see figure 1) and with focal distance of L=133 mm was chosen to verify the method. Additional considerations included practical limitations set by the precision of the tools and raw materials available. As noted in Sec III it was decided to limit the lens diameter to D = 60 mm, based primarily on the estimation of the ability to cut and accurately mount rings for the lens. Figure 4 serves as a guide to the relative location of these parameters.

A compass cutter was used to cut rings from 0.65 mm-thickness flat plastic plate using Vernier calipers to set and verify the width of the cutter. Rings were glued to three supporting posts to form the lens, which was then affixed to the end of an acrylic tube (60 mm diameter, 90 mm length). A baffled circular planar 5-mm-diameter cellular polypropylene polymer (cellular PPP) [33] receiver (KGK, Kawasaki, Japan) was centered 43 mm inside of a second acrylic tube of the same dimension. The use of two tubes allowed for fine-scale adjustment of the lens relative to the receiver, as described below. The scattering target consisted of a fragment of 24 AWG (0.5 mm-diameter) steel wire having a length of approximately 0.62 mm. These dimensions reflect the smallest object that we were able to permanently affix the fragment to a 0.15-mm-diameter cotton thread used for mounting (represented in exaggerated scale as T in figure 4, with the string perpendicular to the figure plane).

The target, lens and receiver were all affixed to a stepper-motor positioning system (LTS-50 Stage and Mark 12 Controller, Sigma Koki, Tokyo, Japan). The target was placed at the Lens focal point, 133 mm from the lens face and the receiver was centered about the opposite focal position. All components of the setup were mounted to facilitate fine adjustment and alignment. In one set of measurements a second thread containing no steel target was also used in order to test for any artifacts that might be due to the mounting thread.

*B. Field Measurements*

The system was aligned and tested using a 400 kHz, 10-cm-diameter, 10-cm-radius-of-curvature focused air transducer (custom-design, KGK, Kawasaki, Japan). Initial alignment entailed situating the target in an arbitrary position approximately on-axis and 1 cm behind the transducer's geometric center. A 30-cycle burst at 400 kHz was used to produce a signal that was received by the system to make small adjustments to the position of the target, the lens and the receiver. Based on the amplitude of the steady-state (center) portion of the received waveform, components were manually adjusted in effort to maximize the signal. It is noted that this alignment-sensitive system required some effort before all components were correctly aligned, with correct positioning evidenced by a sharp position-sensitive rise in amplitude. All components were then rigidly affixed in this position allowing them to move uniformly with the positioning system.

The approximate location of the transducer's geometric focal point was identified by physical measurement and verified by time-of-flight measurement. With the target placed in this point, the positioning system was used to scan radially a distance of 14 mm through the transducer axis of symmetry at spatial intervals of 0.25 mm. Two radial lines separated by $\pi/2$ rad were acquired by rotating the transducer and scanning along a fixed axis. An additional radial measurement was likewise acquired but without a scattering target to determine the effect of the mounting thread.

For all measurements, a 30-cycle 400 kHz burst was generated using an arbitrary waveform generator (NF 1966, Yokohama, Japan) and voltage amplifier (NF HSA 4015, Yokohama, Japan) producing a 200 Vpp input. The maximum voltage response from the receiver was filtered and amplified before being measured with an oscilloscope (LeCroy WaveRunner 6051A, Chestnut Ridge, New York). At each scan point, the voltage amplitude from the oscilloscope was acquired over 100000 points at a sampling rate of 5 GS/s.

Temporal Fourier transforms of the data were used to identify the amplitude and phases of the fundamental, second, and third harmonics of the driving frequency. Field behavior over the focal region was then estimated using a Hankel-transform-based [14] version of previously-described forward and backward planar projection technique [34]. As the projection was sensitive to the input pressure, the 400kHz component of the measured field was normalized to the peak pressure as predicted by the field simulation model (SPL 160 dB). All other relevant values were identical to those used in the field simulations, as provided in Sec III.D predicted to Assuming radial symmetry, the mean of the two radial measurements was used as input, allowing visualization of the field as a function of axial and radial distance.

The amplitude measured along the radial axis through the focus are plotted in Figure 8a for the two radial lines perpendicular to each other, along with their expected pressure distribution. The phase of these measurements are plotted individually in Figures 8b, and 8c, showing good agreement over the central lobe and first three side lobes, but with increased difference in the higher order lobes. Mean percent differences of 10% (green) and 12% (blue) were found in the over 3dB full-width-at-half-maximum (FWHM) and 12% and 17%, respectively over 6dB FWHM.

Projections of the averaged measurements are similarly given in Figure 9. Examination of the Fourier transforms of the time history taken at signal peaks revealed the predicted cancellation of even harmonics and focusing of odd harmonics. An example of the transformed data is shown in Figure 10, where a reduction in signal is apparent near 800 kHz and a small signal peak is visible near 1200 kHz.

## V. Discussion

The present work introduced a binary lens design that possesses the interesting property of focusing both planar waves as well as spherically-diverging waves placed at a focal point. We propose the use of such a lens in a device for charactering fields of small wavelength and/or high intensities of a specific frequency. This concept is demonstrated by constructing a system to measure 400kHz fields in air and performing measurements on a highly-focused air transducer.

Although functionality of a binary-lens-based measurement device was demonstrated, a complete assessment of the method was precluded by the lack of suitable method of independent measurement and, as such was compared with data from simulations. Full evaluation of the concept will require direct comparison of full field measurements against, for example, microphone or optical measurement. These measurement could, for example, be more readily obtained by construction of a similar system tuned to a lower frequency.

The present prototype system was limited in a number of respects that could benefit from more controlled construction. Key limiting factors include the present target size (max dimensions. 0.7 wavelengths), and the diameter of the thread that mounted this target (0.3 wavelengths). Measurements of the thread alone suggest it may have contributed to up to 24% of the overall pressure signal, in general agreement with numeric predictions made over the lens face (17% to 33%). Although potential error introduced by this thread is comparable to that of many ultrasonic receivers [35], a smaller target mounted in a less obtrusive manner would expectedly result in reduced error, at the expense of reduced signal strength. The present study also assumed any motion of the target due to acoustic radiation pressure or streaming was negligible.

Particularly in the characterization of high intensity fields in air, such motion might introduce appreciable error and, if intense enough, would motivate the use of more rigid mounting materials.

An alternative design, for example, might consist of a metal target deposited onto a thin mesh or membrane. Alternatively, the point-like target could be replaced altogether by a thin wire line target whose scattered signals would then be tomographically reconstructed [11]. This would require rotation about the axis of symmetry for full reconstruction; however, it would also simplify the lens design, which would be reduced to a series of linear slits focused onto a rectangular receiver face.

Other design improvements could be implemented for greater signal strength. The most direct change would simply be the use of a larger diameter lens. Consideration of frequency dependence on absorption – an effect not considered in the present work – may also favor a decreased focal length. Such consideration would expectedly be particularly relevant if data from higher (odd) harmonics is desired. Finally, a series of lenses might be incorporated into a design to allow for the imaging of broadband signals. Likewise, study into the effective bandwidth of a single lens, not considered in the present design, is planned for future study.

## VI. Summary and conclusion

A binary lens was used to provide a method for the characterization of a high frequency air HIFU transducer by exploiting the lens' ability to re-focus a spherically spreading wave onto a localized planar region of nearly uniform phase. A system was designed where a scattering target was centered in one focal plane of the lens and a planar receiver centered at the opposing plane. Optimal dimensions were selected through a series of simulations using parameters for measuring 400 kHz signals in air with a 5-mm-diameter receiver placed outside the measurement field. A rudimentary characterization system consisting of a coupled target lens and receiver affixed to a stepping-motor-driven positioning system. Measurements acquired with the prototype system were found to be in overall general agreement with ideal field calculations.

## VII. Acknowledgements

This work was supported by The Center for Industrial and Governmental Relations, The University of Electro-Communications. Scattering and field simulation algorithms utilized in the study were developed under award number R01EB014296 from the National Institute of Biomedical Imaging and Bioengineering of the National Institutes of Health.

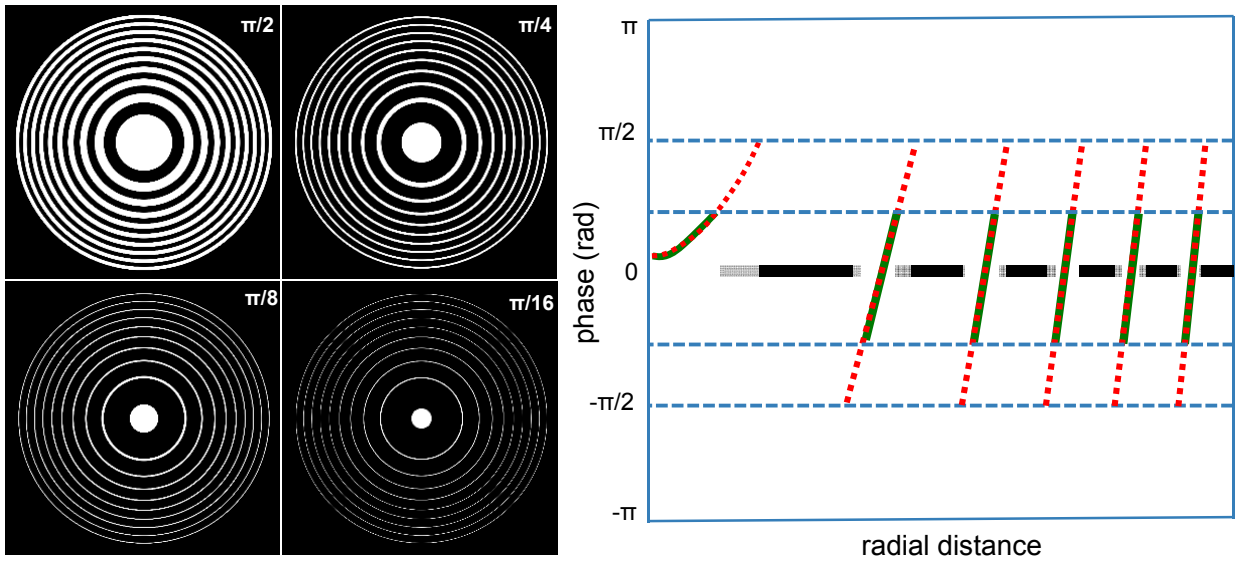

Fig 1

Figure 1: Standard zone plates (Top Left) pass ± π/2 phase variation (Right, dotted). Reducing the slit size about the mid-point of the open rings reduces the variation (solid black to solid grey), producing a bi-convex-like lens behavior at select frequencies.

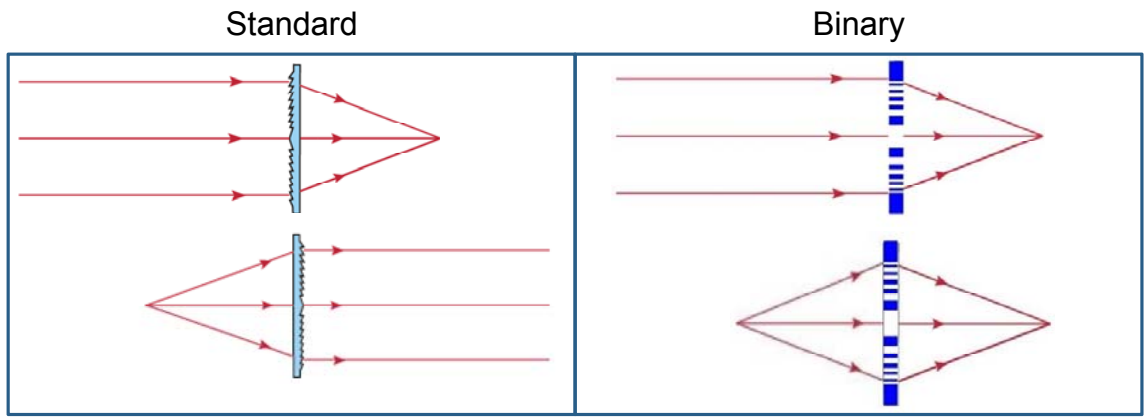

Fig 2

Figure 2. A simple converging-lens (left) focuses a planar wave to a point in the focal plane or, conversely transforms a point source in this plane to a planar wave. The current design (right) focuses a planar wave to a point, and also a point to a point.

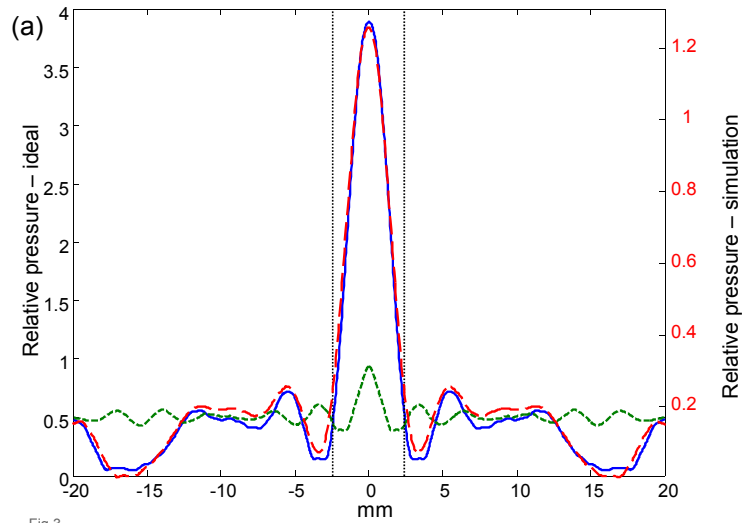

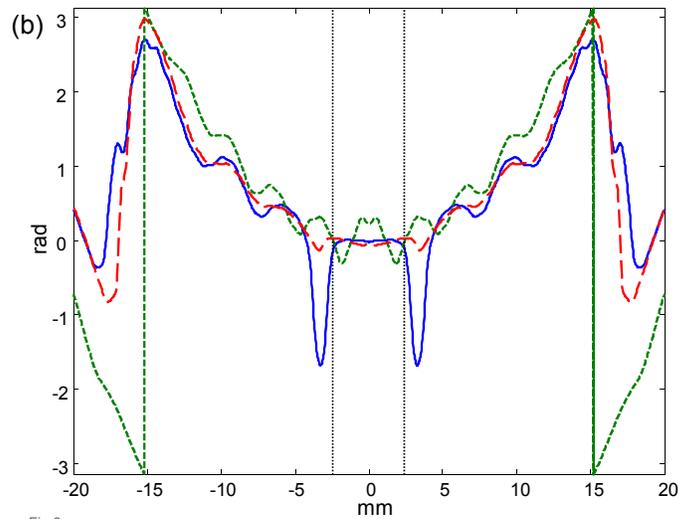

Figure 3. (a) Ideal pressure in the lens focal plane (solid/blue) and the pressure in the same plane if not lens were present (short dash/green). A similar but attenuated profile is predicted in nonlinear simulated data that include effects of attenuation and finite target size (long dash/red). (b) Phase of the same data, illustrating flat phase over the focal width for both the ideal and simulated cases.

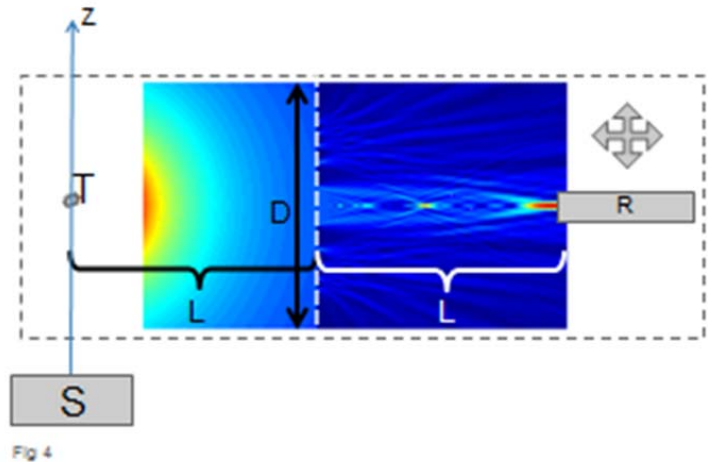

Figure 4. The lens (L), target (T), and receiver (R), are all coupled to move in unison relative to the stationary source transducer (S) transmitting a signal along the z-axis. For field characterization, the uniform motion of (L,R,T) is performed by a stepping-motor-controlled positioner.

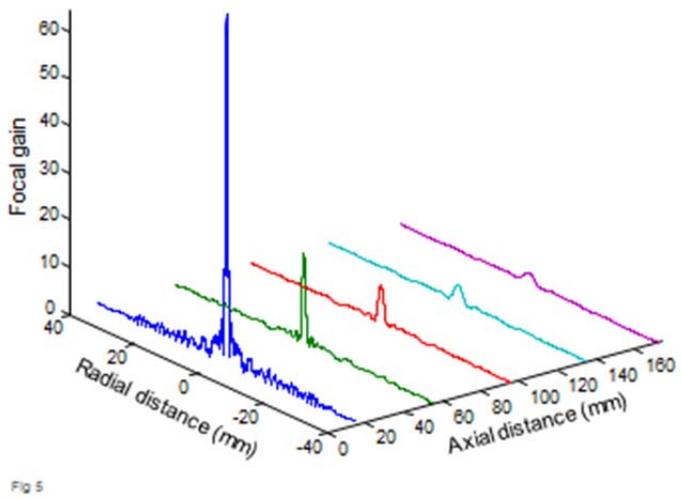

Figure 5. Radial pressure gain in the focal plane for five focal distances.

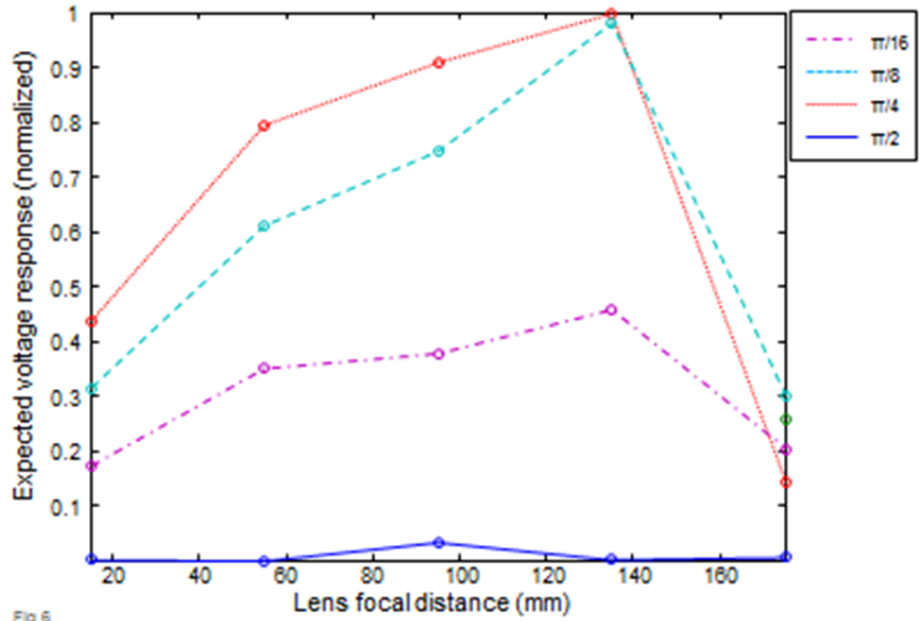

Figure 6. Receiver response vs. focal distance for four different ring configurations. Receiver area covers the first signal peak and response is calculated assuming receiver area Response is assumed linearly proportional to both surface area and mean pressure over the surface.

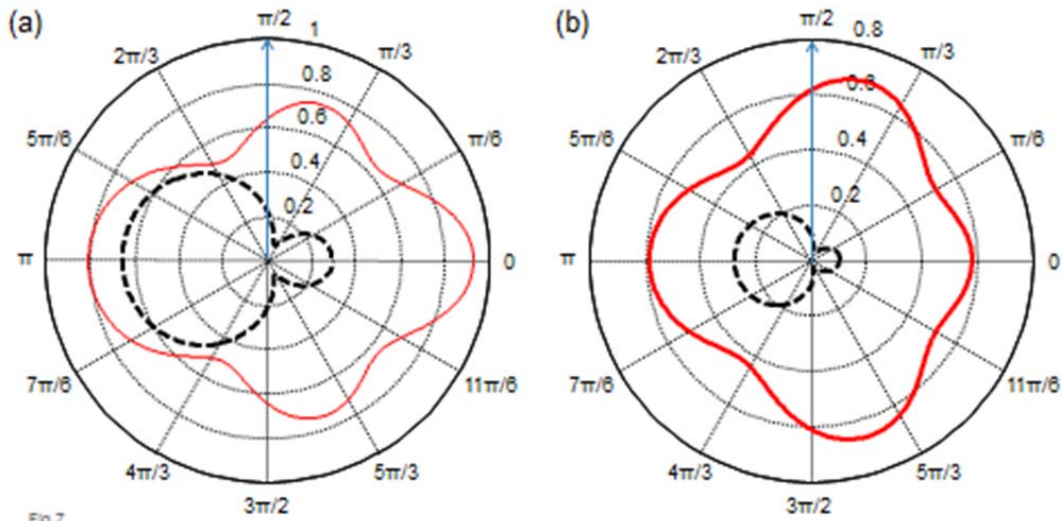

Figure 7. Linear scale pressure scattering profiles of the cylindrical target (solid) and mounting thread (dashed) at (a) z = 0 and (b) z= 60 mm.

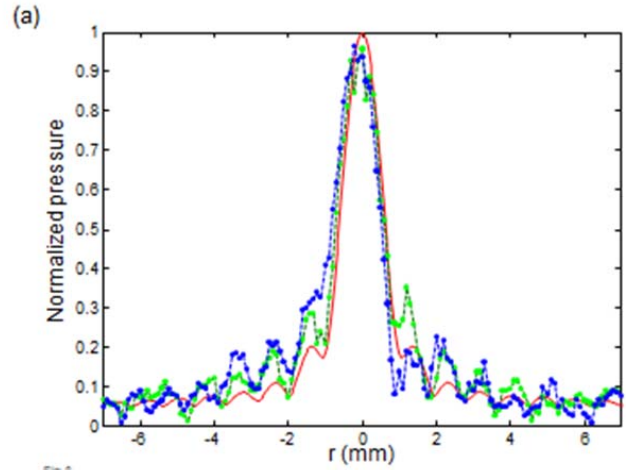

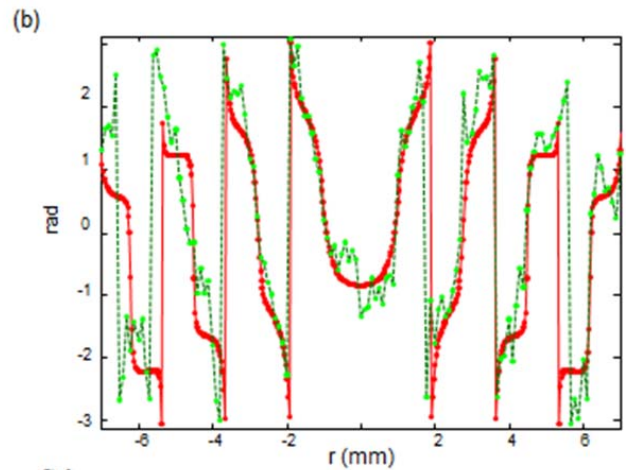

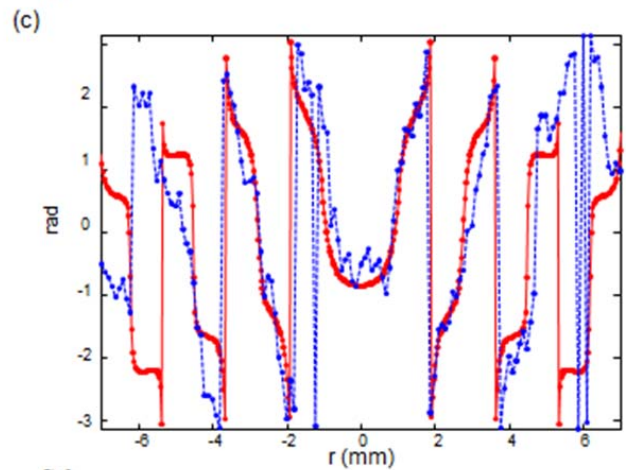

Figure 8. (a). Radial pressure measurements near the focus of a 400 kHz transducer in air. Two radial measurements take perpendicular to each other (dotted and dashed) are compared with a numeric result (solid). (b) Phase of the dotted measurement in (a) compared with numeric result. (c) Phase of the dashed measurement in (a) compared with numeric result.

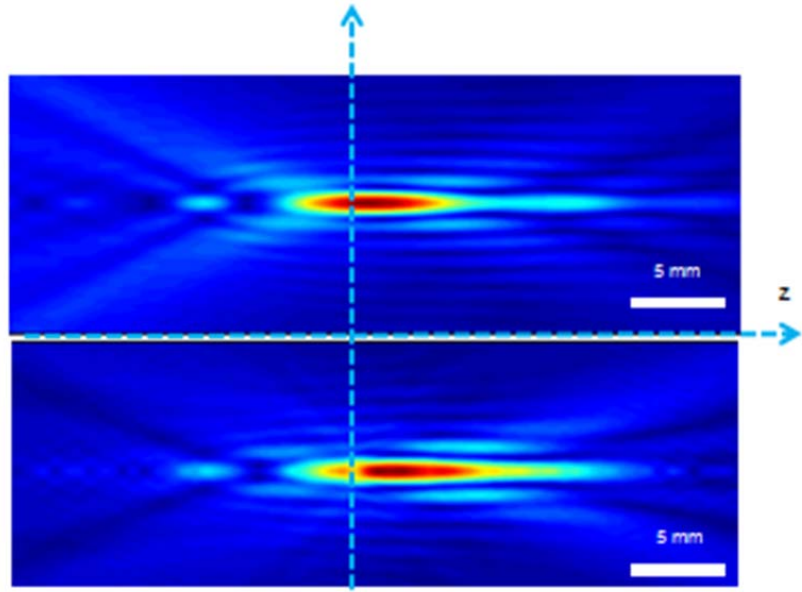

Fig 9

Figure 9. (a) Nonlinear simulation of the 1$^{st}$ harmonic, along the axis of symmetry and (bottom) experimental measurements shown in Figure 8, projected over this same region using a nonlinear forward/backward projection algorithm.

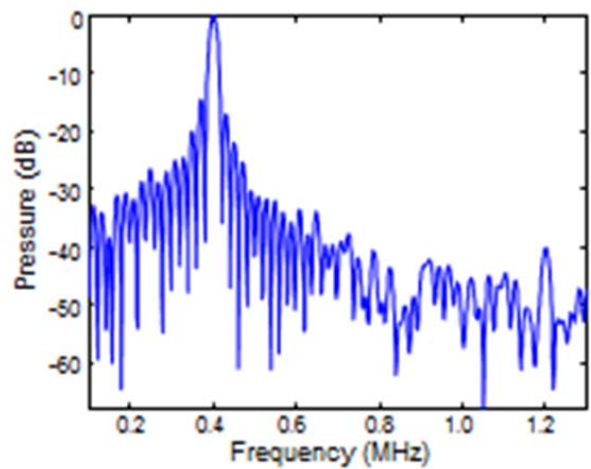

Fig 10

Figure 10. Spectrum of signal near focal peak illustrates cancellation near 0.8 MHz (2$^{nd}$ harmonic) and amplification near 1.2 MHz (3$^{rd}$ harmonic).